\documentstyle[preprint,aps]{revtex}

\begin{document}
\draft
\title{Phase diagram of UPt$_3$ in the $E_{1g}$ model}
\author{K. A. Park and Robert Joynt}
\address{
Department of Physics and Applied Superconductivity Center\\
University of Wisconsin-Madison\\
1150 University Avenue\\
Madison, Wisconsin 53706\\}
\date{\today}
\maketitle
\begin{abstract}
The phase diagram of the unconventional superconductor UPt$_3$
is explained under the long-standing hypothesis that the pair wavefunction
belongs to the $E_{1g}$ representation of the point group.
The main objection to this theory has been that
it disagrees with the experimental phase diagram
when a field is applied along the c-axis.  By a careful
analysis of the free energy this objection is shown to
be incorrect.  This singlet theory also explains the unusual
anisotropy in the upper critical field curves, often thought
to indicate a triplet pair function.
\end{abstract}
\narrowtext

Unconventional superconductivity is a state of matter
under intense discussion at the present time, in both
high-$T_c$ materials \cite{annett} and
the older heavy fermion superconductors.
In this latter class of materials
the most studied and best characterized is UPt$_3$.
The d-wave $E_{1g}$
state was originally proposed as the pairing symmetry on
microscopic grounds \cite{putikka} and has had good success in explaining
a large class of experiments.  It posits a two-component
gap function (in contrast to the d-wave states believed to
be relevant to high-$T_c$) which transforms as ($k_x k_z$,
$k_y k_z$) with corresponding line nodes where the
Fermi surface intersects the plane $k_z=0$ and point nodes
where it intersects the line $k_x=k_y$.  Evidence for this specific
pattern of nodes comes from ultrasound \cite{shivaram} and
heat conduction \cite{benoit} experiments.  $E_{1g}$ also
explains the pressure dependence of the phase diagram. \cite{pressure}
$E_{1g}$, along with other two-dimensional representations of
$D_{6h}$ has a two-component order parameter (OP).  This
leads to a number of unusual predictions which have been
confirmed by experiment, for example the split transition
in specific heat measurements \cite{fisher} and the kink in the
lower critical field curve\cite{hess}.  A two-component OP is
usually (though not always \cite{machida}) accepted for UPt$_3$.

In spite of the fact that $E_{1g}$ has the proper nodal structure
and number of components, a number of alternatives have been proposed.
The objection usually given is that $E_{1g}$
cannot explain the observed phase diagram in the field-temperature
($H$-$T$) plane when $\bbox{H}$ is along the c-axis \cite{garg}.
A second objection
to $E_{1g}$, a singlet theory, is that the
upper critical field curve $H_{c2x}(T)$ for $\bbox{H}$
in the basal plane crosses the curve $H_{c2z}(T)$ for
$\bbox{H}$ perpendicular to the basal plane \cite{cross} and that
this is characteristic of triplet theories.\cite{choi}
In this letter we show that both objections are
unfounded.

It has been clear for several years that the $E_{1g}$ theory
predicts correctly the exceedingly unusual phase diagram in
the $H$-$T$ plane when $\bbox{H}$ is in the basal plane.\cite{joynt}
There are three superconducting phases meeting the normal
phase at a tetracritical point.  Two of these,
the A and C phases, are conventional distorted
Abrikosov lattices formed by one of the two components of the
OP.   The third, the B phase, consists of two interpenetrating
lattices, one formed by each component.  The phases are
separated by second order phase boundaries whose
properties (such as the specific heat jump $\Delta C_V$)
may be calculated.  These conclusions and the conclusions of the
present paper follow from the free energy
density for the $E_{1g}$ theory:
\begin{eqnarray}
\label{fulleng}
f &=& \alpha_0 (T-T_x) |\eta_x|^2 + \alpha_0 (T-T_y) |\eta_y|^2 +
    \beta_1 ( \bbox{\eta} \cdot \bbox{\eta}^*)^2 +
    \beta_2 | \bbox{\eta} \cdot \bbox{\eta}|^2  \\
  & & \mbox{} + \sum_{i,j=x,y} ( K_1 D_i \eta_j D_i^* \eta_j^* +
  K_2 D_i \eta_i D_j^* \eta_j^* + K_3 D_i \eta_j D_j^* \eta_i^*) +
  K_4 \sum_{i=x,y} |D_z \eta_i|^2  \nonumber\\
  & & \mbox{} + (\alpha_0 \epsilon \Delta T) ( \hbar c/2e )
  \sum_{i=x,y} (|D_i \eta_x|^2 -|D_i \eta_y|^2) \nonumber \\
  & & \mbox{} + a_z H_z^2 \bbox{\eta}\cdot\bbox{\eta}^*
  +a_x (H_x^2+H_y^2) \bbox{\eta}\cdot\bbox{\eta}^*
  +a_d |\bbox{H}\cdot\bbox{\eta}|^2.
  \nonumber
\end{eqnarray}
Here $\bbox{\eta} = (\eta_x,\eta_y)$ is the two-component order parameter,
and $K_1$, $K_2$, $K_3$, $K_4$, $\alpha_0$, $\beta_1$, $\beta_2$, $a_x$,
$a_z$, $a_d$ and
$\epsilon$ are constants and $\Delta T = T_x - T_y$. The $D$'s are momentum
operators: $D_x = -i \partial / \partial x + (2e/\hbar c) A_x $ and similarly
for $D_y$ and $D_z$. Here $\bbox{A}$ is the vector potential and $-e$ is
the charge on an electron.
The coupling
of the staggered magnetization to $\bbox{\eta}$ is responsible for the
temperature splitting $\Delta T$.
The existence and need for the term proportional to $\epsilon$, which
represents
the coupling of the supercurrent to the staggered magnetization, and
the terms proportional to $H^2$
was first stressed in the context of a threee-component model.
\cite{m2}

To obtain $H_{c2}$ we need only consider the terms quadratic in
$\bbox{\eta}$ in Eq.\ (\ref{fulleng}).  We minimize these terms following
the Euler-Lagrange perscription of $\delta F = 0$ where $F = \int d^{\/3}x \/
f$.
When the field is in the basal plane this procedure decouples
into one d.e. for $\eta_x$ and one d.e. for
$\eta_y$. Hence we obtain two separate equations for $H_{c2}$.
For appropriate values of the constants these two curves will cross
creating the well-known kink in the upper critical field curve. Hence the A
and C phases correspond to $\bbox{\eta} \sim (1,0)$ and $\bbox{\eta} \sim
(0,1)$ in our theory.  The d.e.'s for $\eta_x$ and $\eta_y$ in this case
both have the same form as the corresponding equation in the more familiar
single component Ginzburg-Landau theory. Hence in
either phase the component of $\bbox{\eta}$ that is non-zero will form an
Abrikosov lattice of the usual kind.  The lattice will be distorted however
because of anisotropy in the gradient terms due to the inequality of the
$K$'s.
In the B phase we have
two flux lattices: one formed by $\eta_x$ and the other formed by
$\eta_y$.\cite{joynt}  The zeroes of these flux lattices need not coincide,
however.
Hence, we must introduce the extra degree of freedom of the offset vector of
the two flux lattices, and we must minimize the free energy with respect to
this variable.  We find that
the free energy is minimized for an offset vector which is one-half of
a flux lattice basis vector.
\cite{prbupt3}

When $\bbox{H}$ is in the $z$-direction, the problem of minimizing the free
energy is far more difficult to solve.  The eigenfunctions of the
linear problem, which we shall call $\bbox{\phi}_{nk}$, can be found
numerically but they are complicated.  Because the linear $H_{c2}$
equations do not separate into separate equations
for $\eta_x$ and $\eta_y$, the
$\bbox{\phi}_{nk}$ have both $x$ and $y$ components.
When the OP expanded in terms of the
eigenfunctions:
$
\bbox{\eta}=\sum_{nk}c_{nk}\bbox{\phi}_{nk},
$
the free energy $F$ is a polynomial in the coefficients:
$F = F(c_{nk})$.  Here $n$ is a level index
(no longer a Landau level index)
and $k$ is the momentum in the $y$-direction.
Part of the argument against $E_{1g}$ runs as follows.
At $H_{c2}$, some of the $c_{0k}$ become nonzero.
The fourth order terms in $F$ are complicated.  If we take, for example,
$\bbox{\eta}=c_{0k}\bbox{\phi}_{0k}+c_{1k}\bbox{\phi}_{1k}$, then
terms of the form $\mid c_{0k}^{} \mid ^2 c_{0k}^* c_{1k}^{}$
appear.  The $c_{0k}$ produce, effectively, a linear term
in the $c_{1k}$.  It is then concluded that no second transition
occurs below $H_{c2}$ in this theory, in conflict with
experiment.

This argument is, however, not correct.  The actual Hilbert space
for the functional $F$ is infinite-dimensional and a
careful analysis of all the possibilities must be carried
out. The energy levels labelled by the integer $n \geq 0$ are highly
degenerate, the eigenvalues being independent of
$k$, which may take the value of any allowed wavevector
$k=2\pi\times \mbox{integer} / L_y$, where $L_y$ is the length of the
sample in the $y$-direction.  The energy of the OP configuration
represented by $\bbox{\phi}_{nk}$ is independent of $k$.
The minimization of $F$ leads to some of the $c_{0k}$ becoming nonzero
at $H_{c2}$ with the formation of the usual hexagonal lattice:
$
c_{nk} \sim \delta_{n,0}(H_{c2}-H)^{1/2} {\cal C}_{k}.
$
Let $2\pi/q$ be the periodicity of the flux lattice in the y
direction.  Then ${\cal C}_{k}$=0 unless $k=mq$, where
$m$ is an integer. As usual ${\cal C}_k  =1(i)$ for $m$ = even(odd).
A dangerous fourth order term in $F$ has the
form:
$
\beta^{01}_{k_1k_2k_3k_4}c_{0k_1}^{*}c_{0k_2}^{}c_{0k_3}^{*}c_{1k_4}^{}.
$
Momentum conservation implies that the coefficient
$\beta^{01}_{k_1k_2k_3k_4}$ is only nonzero if
$k_1+k_2+k_3+k_4=0$.  For an interpenetrating lattice
where the offset vector is one-half of a flux lattice basis vector,
$k_1$, $k_2$, and $k_3$ are integer multiples of $q$, whereas
$k_4$ is half an odd integer times $q$.  Thus the $k$'s
never sum to zero and all dangerous terms vanish.
The $c_{1k}$ for the second lattice never appear in
first order or, by a similar argument, in third order.
{\em The second lattice appears by a second order
transition in the $E_{1g}$ theory for all directions
of the applied field.}  This is in agreement with
experiment and in conflict with previous theoretical
conventional wisdom.  The transition breaks the flux lattice
symmetry because the lattice now has a basis.

We have plotted the phase boundaries obtained by minimizing the
free energy in the following approximation.  The eigenvalues
of the linear $H_{c2}$ operator are obtained by a truncation
of the infinite matrix.  The lowest eigenvalue, which is a function
of $H$ and $T$, gives the $H_{c2}$ curve.  The next lowest eigenvalue
gives a bare inner transition line.  This must be corrected
by an effective field term because the existing lattice
lowers the transition temperature of the new one.
This correction involves only one coupling constant which is
obtained by fitting to the data.
The result is shown in Fig.\ \ref{zplot}.  We have not attempted to fit
the data for $T < 0.4 T_c$ since the linear temperature
dependence of the first two terms in Eq.\ (\ref{fulleng}) breaks down
there.

There is no tetracritical point for
$\bbox{H} = H\hat{z}$; this is due to level repulsion.
We regard this as a virtue of the theory, because the experimental data
show that to call the phase diagram isotropic is an exaggeration.
The $H_{c2}$ curve for $\bbox{H} = H \hat{z}$ does not have a kink,
only a flat region well reproduced by the
theory, and the data are consistent with only two superconducting phases,
as the present theory predicts for this field direction.
Another part of the argument against E$_{1g}$ has been that fine tuning
of parameters is required to fit the data.  Our fit does not
does not require any fine-tuning.  We find that the
phase diagram for {\em both} field directions can be
fit by the same set of parameters, and the only numerical coincidence
which arises when this is done is that $K_2 \approx K_3$.
This is actually a consequence of approximate
particle-hole symmetry and the fact
that it comes out of the fit is further evidence that the overall
picture is correct.

To understand the directional dependence of $H_{c2}$,
it is first necessary to discuss the
magnetic susceptibility of UPt$_{3}$.
This issue is complicated by the fact that all renormalizations involved
are not well understood.  Since UPt$_3$ is a Fermi liquid, however,
the starting point must be the single-particle states calculated in
band theory, which accounts very well for the
Fermi surface. \cite{wang}  The
states near the Fermi surface are predominantly derived from
uranium 5f orbitals with $j=5/2$.
In the isolated atom, these would be
6-fold degenerate.  In the hexagonal crystal field,
there is an effective Hamiltonian of the form
$H_{crystal field}=B_{h}(j_{z}^{2}-j(j+1)/3)$,
where $B_{h}$ is a constant.  This splits the six-fold
degenerate state into three doublets at the $\Gamma$ point:
$j_z = \pm 5/2$, $j_z = \pm 3/2$, and $j_z = \pm 1/2$.  There is also
an even-odd splitting from the fact that there are two U atoms per unit cell.
The six bands constructed from these
states cross the Fermi energy, and the crystal field splitting is of the
same order as the bandwidth.  The average occupation of the
5f level is between 2 and 3.  If we apply a magnetic field, there will be both
a Pauli (intraband) and a Van Vleck (interband) contribution to the
susceptibility.  The former is of order $(g_{\mbox{\scriptsize \em eff}}
 \mu_B)^2 N(\varepsilon_F)$,
while the latter is of order $(g_{\mbox{\scriptsize \em eff}} \mu_B)^2/
 \mid B_h \mid$.  Here $g_{\mbox{\scriptsize \em eff}}$ is
an effective g-factor for the coupling of the field to the total
angular momentum of the band or bands involved.  It is a dimensionless
number of order unity.  The Land\'{e} factor for $\ell$=3, s=1/2 ,
and $j$=5/2 is 6/7.

The Van Vleck susceptibility is given by
\begin{equation}
\chi_{ii}=2n\mu_{B}^{2} \sum_{\alpha,\beta}
\frac
{|<\alpha|L_{i}+2 S_{i}|\beta>|^2}
{E_{\beta}-E_{\alpha}}
f_{\alpha}(1-f_{\beta}).
\end{equation}
Here $f_\alpha$, $f_\beta$, $E_\alpha$, $E_\beta$ are occupation factors
and energies of the states $\alpha$ and $\beta$.
In view of the greater
multiplicity of the interband transitions, we
expect the Van Vleck susceptibility to be very important - indeed it
may dominate the total.
If $\bbox{H}$ is along the c-axis, then
the relevant matrix element (with $\hbar = 1$) is:
\begin{equation}
|<\alpha| L_z+2 S_z|\beta>|^2
=(36/49)j_z^2 \delta_{\alpha,\beta}.
\end{equation}
In the approximation that states of different $j_z$
do not mix (negligible intersite interactions),
then the perturbation introduced by $\bbox{H}$ is diagonal, and
the occupation factors then imply that the Van Vleck susceptibility is
zero for this direction.  If $\bbox{H}$ is in the $x$-direction,
the corresponding expression for the square of the matrix element is
\begin{equation}
|<\alpha| L_x+2 S_x|\beta>|^2 =
(36/49)(5/2-j_z)(5/2+j_z+1)
\end{equation}
if the states $\alpha$ and $\beta$ differ by one unit of $j_z$
and is zero otherwise.  The Van Vleck susceptibility comes from
four distinct pairs of states:($j_z$=-5/2,-3/2), (-3/2,-1/2), (1/2,3/2) and
(3/2,5/2), whenever one of the pair is occupied and the other unoccupied.
The Pauli contribution to $\chi_{xx}$, on the other hand, comes only from
the pair (-1/2,1/2) when this state is occupied.

Summing up these considerations, we expect that $\chi_{zz}$ will be
dominated by the Pauli contribution, while $\chi_{xx}$ will be
dominated by the Van Vleck contribution.  There are two effects which can
vitiate these conclusions.
Intersite effects will mix states of different $j_z$, and this will
modify this ionic picture.  The interaction effects give rise to the
large Fermi liquid enhancement of the
susceptibility, which comes chiefly from the mass term.
This is expected to affect Pauli and Van Vleck terms alike.\cite{fuchun}
Experiment confirms these theoretical arguments.
It is found that $\chi_{xx}$ is considerably larger than $\chi_{zz}$ at
all temperatures, in accord with the expectation that the Van Vleck
contribution is large.  The
temperature dependence of $\chi_{xx}(T)$ is anomalous, with a peak
at T=15 K.\cite{frings}  This peak is absent in the
smooth curve for $\chi_{zz}(T)$, and in the
the specific heat $C_V(T)$. \cite{devisser}
This is consistent with the idea that the physical origins of
$\chi_{zz}$ and $\chi_{xx}$ are different, and that the
density of states at the Fermi level
determines $\chi_{zz}$ but not $\chi_{xx}$.
Thus experiment confirms the theoretical picture.

The importance of these considerations for the superconducting
state is simple.  \cite{cox} Superconductivity affects the Pauli susceptibility
in a drastic fashion.  For a singlet state such as $E_{1g}$,
the Pauli term $\chi_{ij}^P(T)$
is reduced to zero at zero temperature because
it takes a finite amount of energy to break a pair and magnetize
the system.  Superconductivity should have no effect at all
on the Van Vleck term.  The difference in free energies between the
normal and superconducting states in a field is
$
F_{magnetic}=-\frac{1}{2} \sum_{ij} \Delta\chi^P_{ij} H_{i} H_{j}.
$
Here $\Delta \chi_{ij}^P = \chi_{ij}^S - \chi_{ij}^N$ where $\chi_{ij}^S$
and $\chi_{ij}^N$ are the Pauli susceptibilities in the superconducting and
normal states, respectively.
Hence we expect a field along the c-axis to have the largest
effect on superconductivity.
Near $T_c$, the slope of $H_{c2}$, larger in magnitude
for $\bbox{H}$ in the z-direction, is determined by the
terms in $F$ which are linear in $H$.  The
different slopes reflect the
anisotropic coherence length and are not directly related to the
susceptibility.
As $H$ increases, the $H^2$ terms become more important and cause
$H_{c2}(T)$ to curve down.  The anisotropy in the Pauli
susceptibility then causes $H_{c2z}$ to
curve more strongly with the result that the two curves cross.
To implement this quantitatively, we note that the
change in the susceptibility is
quadratic in $\bbox{\eta}$ near $T_c$.  The expression
for $F_{magnetic}$ which results
is precisely the last three terms, proportional to
$H^2$, in Eq.\ (\ref{fulleng}).
The resulting fit is shown in Fig.\ \ref{crossing}.

What these arguments show is that the peculiar anisotropy
of the upper critical field is evidence for
a {\em singlet} superconducting state, such as $E_{1g}$.
This is in sharp contrast to previous arguments that
the anisotropy points to a triplet state.  These arguments
were based on the idea that the observed anisotropy
in the total susceptibility is also reflected in the
Pauli term, that is $\chi^P_{xx} \approx 2 \chi^P_{zz}$.
According to the above arguments, this appears unlikely.

We conclude that the $E_{1g}$ theory can account for
two crucial aspects of the phase diagram of UPt$_3$:
the existence and shape of the inner transition line
for $\bbox{H}$ along the c-axis, and the peculiar
anisotropy of the upper critical field.  This
removes the major objections to this theory, which
otherwise gives a good account of the low temperature
thermodynamics, including the position of the gap nodes,
the tetracritical point, and the dependence of the
phase boundary positions on applied pressure.

We are very grateful to D. L. Cox for useful discussions
and acknowledge the support of the
National Science Foundation through grant no. 9214739.

\begin{figure}
\caption{Phase diagram when the field is in the $z$-direction. The lines
are the theoretical fits to $H_{c2}$ (solid line) and the inner transition
(dashed line).  The data points are from ultrasonic velocity measurements
and are taken from Ref.\ \protect \cite{adenwalla}, Fig.\ 3.}
\label{zplot}
\end{figure}

\begin{figure}
\caption{Shows the crossing of the $H_{c2}$ line when the field is the
basal plane (solid line) and the $H_{c2}$ line when the field is in the
$z$-direction (dashed line).  The data points are from ultrasonic velocity
measurements and are taken from Ref.\ \protect \cite{adenwalla}, Fig.\ 3. The
diamonds are for the case when the field in the basal plane
($\protect \bbox{H} \parallel $ ab)
and the crosses are for the case when the field is in the $z$-direction
($\protect \bbox{H} \parallel $ c).}
\label{crossing}
\end{figure}

\begin{references}
\bibitem{annett} J. F. Annett, N. Goldenfeld, and S. Renn, Phys. Rev. B
{\bf 43}, 2778 (1991); a recent review is by D. Scalapino
(unpublished)
\bibitem{putikka} W. O. Putikka and R. Joynt, Phys. Rev. B{\bf 49},
701 (1989)
\bibitem{shivaram} B. S. Shivaram, Y. Jeong, T. Rosenbaum, and D. Hinks,
Phys. Rev. Lett. {\bf 56}, 1078 (1986);P. Hirschfeld, D. Vollhardt,
and P. W\"{o}lfle, Sol. St. Comm. {\bf 59}, 111 (1986)
\bibitem{benoit} B. Lussier, B. Ellmann, and L. Taillefer (unpublished)
\bibitem{pressure} R. Joynt, Phys. Rev. Lett., {\bf 71},
3015, (1993)
\bibitem{fisher} R. Joynt, Supercon. Sci. Tech. {\bf 1}, 210 (1988);
R. A. Fisher {\em et al.}, Phys. Rev. Lett. {\bf 62},
1441 (1989); K. Hasselbach, L. Taillefer, and J. Flouquet, Phys. Rev. Lett.
{\bf 63}, 93 (1989)
\bibitem{hess}D. W. Hess, T. Tokuyasu, and J. Sauls,
J. Phys. Cond. Matt. {\bf 1}, 8135 (1989); B. S. Shivaram, J. J. Gannon,
and D. Hinks, Phys. Rev. Lett. {\bf 63}, 1441 (1989)
\bibitem{machida} K. Machida and M. Ozaki, Phys. Rev. Lett.
{\bf 66}, 6293 (1991)
\bibitem{garg} See, for example,
A. Garg, Phys. Rev. Lett. {\bf 69}, 676 (1992);
J. A. Sauls, Adv. in Phys. {\bf 43}, 113 (1994)
\bibitem{cross}B. Shivaram, T. Rosenbaum, and D. Hinks,
Phys. Rev. Lett. {\bf 57}, 1259 (1986)
\bibitem{choi} C. Choi and J. Sauls, Phys. Rev. Lett. {\bf 66}, 484 (1991)
\bibitem{joynt} R. Joynt, Europhys. Lett. {\bf 16}, 289 (1991)
\bibitem{m2} K. Machida, T. Ohmi, and M. Ozaki, J. Phys. Soc. Japan
{\bf 62}, 3216 (1993)
\bibitem{prbupt3} K. A. Park and R. Joynt, to be published;
there is a commensuration energy at this offset vector, as in
D.-C.Chen and A. Garg, Phys. Rev. B {\bf 49}, 479 (1994)
\bibitem{wang} C.S. Wang {\em et al.}, Phys. Rev. B {\bf 35},
7260 (1987)
\bibitem{fuchun} F. C. Zhang and T. K. Lee, Phys. Rev. Lett.
{\bf 58}, 2728 (1987); C. Varma and G. Aeppli, {\em ibid.\ }
{\bf 58}, 2729 (1987); D. L. Cox, {\em ibid.\ }
{\bf 58}, 2730 (1987)
\bibitem{frings} P. Frings, J. Franse, F. deBoer, and A. Menovsky,
J. Mag. Mag. Mat., {\bf 31}, 240 (1983)
\bibitem{devisser} A. de Visser, A. Menovsky, and J. Franse,
Physica B, {\bf 147}, 81 (1987)
\bibitem{cox} This point has been stressed particularly
by D. L. Cox (private communication)
\bibitem{adenwalla} S. Adenwalla, S. W. Lin, Q. Z. Ran, Z. Zhao, J. B.
Ketterson, J. A. Sauls, L. Taillefer, D. G. Hinks, M. Levy, and Bimal K.
Sarma, Phys. Rev. Lett. {\bf 65}, 2298 (1990)
\end{references}
\end{document}